\documentclass{svjour3}                     
\smartqed  
\usepackage{graphicx}
\usepackage{epsfig}
\newcommand{\HI}{H\,{\sc i}}

\newcommand{\tel}{ASKAP}

\newcommand{\lesssim}{\mathrel{\hbox{\rlap{\hbox{\lower4pt\hbox{$\sim$}}}\hbox{$<$}}}}
\newcommand{\gtrsim}{\mathrel{\hbox{\rlap{\hbox{\lower4pt\hbox{$\sim$}}}\hbox{$>$}}}}

\journalname{Experimental Astronomy}
\begin{document}

\pagenumbering{roman}

\title{SCIENCE WITH ASKAP
}
\subtitle{The Australian Square-Kilometre-Array Pathfinder}


\author{
S. Johnston$^{*}$\thanks{\bf{$^{*}$Email: Simon.Johnston@csiro.au}} \and R. Taylor \and M. Bailes 
\and N. Bartel \and C. Baugh 
\and M.~Bietenholz \and C. Blake \and R. Braun \and J. Brown \and 
S. Chatterjee \and J. Darling \and A. Deller \and R.~Dodson \and P.~Edwards 
\and R.~Ekers \and  S.~Ellingsen \and I.~Feain \and B.~Gaensler \and M.~Haverkorn 
\and G.~Hobbs \and A.~Hopkins \and C.~Jackson \and C.~James \and G.~Joncas 
\and V.~Kaspi \and V.~Kilborn \and B.~Koribalski \and R.~Kothes
\and T.~Landecker \and A.~Lenc \and J.~Lovell \and J.-P.~Macquart
\and R.~Manchester \and D.~Matthews \and N.~McClure-Griffiths \and R.~Norris
\and U.-L.~Pen \and C.~Phillips \and C.~Power \and R.~Protheroe \and E.~Sadler
\and B.~Schmidt \and I.~Stairs \and L.~Staveley-Smith \and J.~Stil \and  
S.~Tingay \and A.~Tzioumis \and M.~Walker \and J.~Wall$^{\dagger}$\thanks{\bf{$^{\dagger}$Overall Editor}} 
\and M.~Wolleben
}


\institute{S. Johnston \and R. Braun \and P. Edwards \and R. Ekers \and I. Feain \and
G. Hobbs \and C. Jackson \and B. Koribalski \and R. Manchester \and N. McClure-Griffiths 
\and R. Norris \and C. Phillips \and A. Tzioumis \at
   Australia Telescope National Facility, CSIRO, PO Box 76, Epping, NSW 1710, Australia
           \and
R. Taylor \and J. Brown \and R. Kothes \and J. Stil \at
   Dept of Physics and Astronomy, University of Calgary, Calgary, AB T2N 1N4, Canada
\and
M. Bailes \and C. Blake \and A. Deller \and V. Kilborn \and A. Lenc \and C. Power \at
Centre for Astrophysics and Supercomputing, Swinburne University of Technology,
PO Box 218, Hawthorn, Vic 3122, Australia
\and
N. Bartel \and M. Bietenholz \at
Dept of Physics and Astronomy, York University, Toronto, ON M3J 1P3, Canada
\and
C. Baugh \at
Institute for Computational Cosmology, University of Durham, Durham, DH1 3LE, UK
\and
M. Bietenholz \at
Hartebeesthoek Radio Observatory, PO Box 443, Krugersdorp 1740, South Africa
\and
S. Chatterjee \and B. Gaensler \and A. Hopkins \and E. Sadler\at
School of Physics, The University of Sydney, NSW 2006, Australia
\and
J. Darling \at
Center for Astrophysics and Space Astronomy, University of Colorado, 389 UCB, Boulder, CO 80309-0389, USA
\and
R. Dodson \at
Observatorio Astronomico Nacional, Alcara de Henares, Spain
\and
S. Ellingsen \and J. Lovell \at
School of Mathematics and Physics, University of Tasmania, Private Bag 21, Hobart, Tas 7001, Australia
\and M. Haverkorn \at
NRAO Jansky Fellow: Astronomy Dept, University of California-Berkeley, Berkeley, CA 94720, USA
\and 
C. James \and R. Protheroe \at
School of Chemistry \& Physics, University of Adelaide, SA 5006, Australia
\and
G. Joncas \at
Dept de Physique et Observatoire du Mont Megantic, Universite Laval, Quebec, QC G1K 7P4, Canada
\and
V. Kaspi \at
Dept of Physics, McGill Unversity, Montreal, QC H3A 2T8, Canada
\and
T. Landecker \and M. Wolleben \at
Dominion Radio Astrophysical Observatory, Herzberg Institute of Astrophysics, NRC, Penticton, BC, Canada
\and
J.-P. Macquart \at
NRAO Jansky Fellow: Astronomy Dept, California Institute of Technology, Pasadena, CA 91125, USA
\and
D. Matthews \at
Dept of Physics, La Trobe University, Vic 3086, Australia
\and
U.-L. Pen \at
Canadian Insititute for Theoretical Astrophysics, University of Toronto, Toronto, ON M5S 3H8, Canada
\and
B. Schmidt \at
Mount Stromlo and Siding Spring Observatory, Private Bag, Weston Creek, Canberra, ACT 2601, Australia
\and
I. Stairs \and J. Wall \at
Dept of Physics and Astronomy, University of British Columbia, 6224 Agricultural Road, Vancouver, BC V6T 1Z1, Canada
\and
L. Staveley-Smith \at
School of Physics, University of Western Australia, Crawley, WA 6009, Australia
\and
S. Tingay \at
Dept of Imaging and Applied Physics, Curtin University of Technology, Bentley, WA, Australia
\and
M. Walker \at
Manly Astrophysics Workshop Pty Ltd, Manly, NSW 2095, Australia
}
\date{Received: date / Accepted: date}

\maketitle

%
%
\clearpage
\newpage
\section*{Abstract}
The future of cm and m-wave astronomy lies with 
the Square Kilometre Array (SKA), 
a telescope under development by a consortium of 17 countries.
The SKA will be 50
times more sensitive than any existing radio facility. 
A majority of the key science for the SKA will be addressed through large-area
imaging of the Universe at frequencies from 300 MHz to a few GHz.\\

\noindent
The Australian SKA Pathfinder (\tel) is aimed squarely 
in this frequency range, and achieves instantaneous wide-area imaging 
through the development and deployment of phase-array feed systems on 
parabolic reflectors. This large field-of-view makes \tel\ an unprecedented synoptic 
telescope poised to achieve substantial advances in SKA key science.
The central core of \tel\ will be located at the Murchison Radio
Observatory in inland Western Australia, one of 
the most radio-quiet locations on the Earth and one of
the sites selected by the international community as a potential 
location for the SKA.\\

\noindent
Following an introductory description of ASKAP, this document contains 
7 chapters describing specific science programmes 
for \tel. In summary, the goals of these programmes are as follows:
\begin{itemize}
\item The detection of a million galaxies in atomic hydrogen emission 
across 75\% of the sky out to a redshift of 0.2 to understand galaxy 
formation and gas evolution in the nearby Universe.
\item The detection of synchrotron radiation from 60 million galaxies to 
determine the evolution, formation and population of galaxies across 
cosmic time and enabling key cosmological tests.
\item The detection of polarized radiation from over 500,000 galaxies, 
allowing a grid of rotation measures at $10'$  to explore
the evolution of magnetic fields in galaxies over cosmic time.
\item The understanding of the evolution of the interstellar medium of 
our own Galaxy and the processes that drive its chemical and physical 
evolution.
\item The high-resolution imaging of intense, energetic phenomena by 
enlarging the Australian and global Very Long Baseline networks.
\item The discovery and timing of  a thousand new radio pulsars.
\item The characterization of the radio transient sky through 
detection and monitoring of transient sources such as gamma ray 
bursts, radio supernovae and intra-day variables.
\end{itemize}
\noindent
The combination of location, technological innovation and scientific program
will ensure that \tel\ will be a world-leading radio 
astronomy facility, closely aligned with the scientific and 
technical direction of the SKA. A brief summary 
chapter emphasizes the point, and considers discovery space.

\vspace*{10mm}
This astro-ph submission contains only an outline of the entire
document published by Experimental Astronomy. You can download
the full article from\\
http://wwwatnf.atnf.csiro.au/projects/askap/newdocs/askap\_expast08.pdf

\newpage
\section{Introduction}\label{chap:intro}
\pagenumbering{arabic}
\setcounter{page}{1}
{\bf Section authors: {\em S. Johnston, R. Taylor}}\\

The Australian SKA Pathfinder (\tel) is a next generation radio 
telescope on the strategic
pathway towards the staged development of the Square Kilometre Array 
(SKA; see Schilizzi et al. 2007 for preliminary SKA specifications). 
The \tel\ project is international in scope and includes 
partners in Australia, Canada, the Netherlands and South Africa.
This document, which concentrates on the science made possible with \tel\
was written as a joint collaboration between Australian and Canadian
research scientists.
\\

\noindent \tel\ has three main goals:
\begin{itemize}
\item to carry out world-class, ground breaking observations directly
relevant to the SKA Key Science Projects,
\item to demonstrate and prototype the technologies for the mid-frequency
SKA, including field-of-view enhancement  by focal-plane phased arrays
on new-technology 12-metre class parabolic reflectors,
\item to establish a site for radio astronomy in Western Australia where 
observations can be carried out free from the harmful effects of 
radio interference.
\end{itemize}

\tel\ is part of the Australian strategic pathway towards the SKA as outlined
in the Australian SKA Consortium Committee's ``SKA: A Road Map for
Australia'' document. \tel\ is seen as `...a significant scientific
facility, maintaining Australia's leading role within the SKA
partnership and addressing key outstanding computational/calibration
risk areas ...'. SKA programs were given the highest priority for
Australian radio astronomy in the 2006$-$2015 Decadal Plan for Astronomy.

In Canada the partnership in development and
construction of \tel\ forms part of the Canadian SKA program, 
funded as part of one of the top priorities for future facilities in 
the Canadian Long Range Plan for Astronomy.
In November 2006 the President of National Research Council of Canada and 
the Chief Executive Officer of the CSIRO signed an agreement declaring their 
intention to collaborate in the realization of \tel.

\newpage
\section{Extragalactic \HI\ Science}\label{chap:hi}
{\bf Lead authors: {\em L. Staveley-Smith, U.-L. Pen}}\\
{\bf Contributing authors: {\em C. Baugh, C. Blake, R. Braun, J. Darling,
V. Kilborn, B. Koribalski, C. Power, E. Sadler}}\\

\subsection{Summary}
Understanding how galaxies form and evolve is one of the key
astrophysical problems for the 21st century. Since neutral hydrogen (\HI) is
a fundamental component in the formation of galaxies, being able to
observe and model this component is important in
achieving a deeper understanding of galaxy formation. Widefield \HI\
surveys using the next generation radio telescopes such as \tel\ and
ultimately the SKA will allow unprecedented insights into the
evolution of the abundance and distribution of neutral hydrogen with
cosmic time, and its consequences for the cosmic star formation, the
structure of galaxies and the Intergalactic Medium. \tel\ will
provide powerful tests of theoretical galaxy formation models and
improve our understanding of the physical processes that shaped the
galaxy population over the last $\sim$7 billion years.

\newpage
\section{Continuum Science}\label{chap:cont}
{\bf Lead authors: {\em I. Feain, J. Wall}}\\
{\bf Contributing authors: {\em C. Blake, R. Ekers, A. Hopkins, C. Jackson, R. Norris}}\\

\subsection{Summary}
Understanding the formation and evolution of galaxies and active galactic
nuclei (AGN) as a function of cosmic time is one of the most intensely
investigated issues in contemporary astronomy, and is a key science driver
for next-generation telescopes such as SKA, ALMA and ELTs. Today's
instruments already give profound insights into the galaxy population at
high redshifts, and a number of current surveys (e.g. ATLAS, ELAIS, GOODS,
COSMOS) are asking questions such as: When did most stars form? How do AGNs
influence star and galaxy formation? What is the spatial distribution of evolved
galaxies, starbursts, and AGNs at $0.5 < z < 3$? Are massive black holes a
cause or a consequence of galaxy formation?

However, most of these surveys are primarily at optical and IR wavelengths,
and can be significantly misled by dust extinction. Radio observations
not only overcome dust extinction, but also provide data on AGN that
are unavailable at other wavelengths. \tel\ will be able to determine how
galaxies formed and evolved through cosmic time, by penetrating the
heavy dust extinction which is found in AGN at all redshifts, and
studying the star formation activity and AGN buried within.

\newpage
\section{Polarization Science}\label{chap:poln}
{\bf Lead authors: {\em R. Taylor, B. Gaensler}}\\
{\bf Contributing authors: {\em J. Brown, M. Haverkorn, R. Kothes, T. Landecker,
J. Stil, M. Wolleben}}\\

\subsection{Summary}
\tel\ will be both a technical and scientific pathfinder to the Square
Kilometre Array.  One of the five key-science drivers for the SKA is
to understand the origin and evolution of cosmic magnetism.  The
prime observational approaches to this key science will be a deep
survey of polarized emission from compact extragalactic sources (Beck
\& Gaensler 2004), and of the diffuse polarized radiation from the
Galaxy.  This data set will provide
\begin{itemize}
\item a deep census of the polarization properties of galaxies as a 
function of redshift (secured through complementary \HI\ or optical surveys),
\item a dense grid of Faraday rotation measures to over 
500,000 background radio sources
\item all-sky Faraday Rotation image of the 3-dimensional structure of the  diffuse 
magneto-ionic medium of the Galaxy.
\end{itemize}

The Rotation Measure (RM) of a polarized radio signal is derived from the variation in 
wavelength of the polarization angle of linearly polarization, $\phi$, given by
\begin{equation}
\phi  =  \phi_\circ + \lambda^2 \; 0.812 \int{n_e}{\bf B}\cdot {\bf
dl} =   \phi_\circ + \lambda^2 \;  {\rm RM} \\
\label{eq-RM}
\end{equation}
where $n_e$ [cm$^{-3}$] is the electron density and {\bf B} [$\mu$G]
is the magnetic field along the propagation path {\bf dl} [pc].
A dense, all-sky grid of RM values to background radio sources and 
all-sky images of the RM of the polarized emission from the Galaxy, are 
thus powerful and unique probes of cosmic magnetism in the Milky Way Galaxy,
the intergalactic medium, and in extragalactic radio sources. 

\newpage
\section{Galactic and Magellanic Science}\label{chap:gal}
{\bf Lead authors: {\em N. McClure-Griffiths, G. Joncas}}\\
{\bf Contributing author: {\em D. Matthews}}\\

\subsection{Summary}
The rebirth of observational studies of the Milky Way and Magellanic
System over the past decade has raised new and profound
questions about the evolution of the interstellar medium (ISM).
The community has transitioned from studying small-scale aspects 
of the ISM to a more wholistic approach, which seeks to combine 
information about a variety
of ISM phases with information about magnetic fields.  With \tel\ we can make
significant and unique inroads into understanding the evolution of the
ISM and through that the evolution of the Milky Way.  These are crucial
steps along the path to understanding the evolution of galaxies.  

The Milky Way and Magellanic System because of their very large sky
coverage can only be observed in survey mode.  As such, they present
themselves as ideal targets for \tel.  Here we discuss how \tel\ can
be used to achieve superb advances in our understanding of the
evolution of the Milky Way, its ISM and its magnetic field. To achieve
these advances we propose several large-scale projects, which include:
accounting for and studying the structure of all \HI\ associated with
the Milky Way and Magellanic System to unprecedented column-density
limits and angular resolution; constraining the large-scale Galactic
magnetic field in the inner Galaxy with {\em in situ} measurements of
\HI\ Zeeman splitting; exploring the growth of molecular clouds using
diffuse OH mapping of the Galactic plane; and probing the turbulent
magneto-ionic medium of the Galactic halo with diffuse continuum
polarization mapping of the whole sky.  The unifying technological
requirement for all aspects of the Galactic and Magellanic science
presented herein is the demand for low-surface-brightness
sensitivities achieved by compact configurations.

\newpage
\section{Very Long Baseline Interferometry Science}\label{chap:vlbi}
{\bf Lead authors: {\em S. Tingay, N. Bartel}}\\ 
{\bf Contributing authors: {\em M. Bietenholz, A. Deller, R. Dodson, P. Edwards, S. Ellingsen, 
A. Lenc, J. Lovell, C. Phillips, A. Tzioumis}}\\

\subsection{Summary}
We describe possible areas of science that can be
addressed using VLBI and \tel.  A number of scientific programs 
present themselves when considering the use of \tel\ 
as part of the Australian Long Baseline Array (LBA) and the 
global VLBI array.  Better angular resolution at L-band, better 
sensitivity, and better {\em uv} coverage will aid standard VLBI observations 
of AGN, pulsars, and OH masers.  An innovative additional capability 
for \tel\ is multibeaming.  If this can be harnessed for VLBI in the form of 
multiple phased array beams, a number of wide-field survey observations become 
feasible.

In time, \tel\ should also become part of the recently 
developed Australian e-VLBI network, currently called PAMHELA.
An interesting possibility is to not use \tel\ as part of the LBA, but use 
it as a source of trigger information for radio transients, as part of
\tel\ survey work.  These triggers could be transmitted to PAMHELA, and the 
candidate sources targeted at high angular resolution in rapid follow-up 
observations.  The combination of \tel\ and 
PAMHELA would be unique and powerful.  PAMHELA would add great scientific 
value to the low resolution detection of transients by \tel.

At this point the VLBI community supports the extension of the \tel\ frequency 
range to S-band (2.4~GHz) but recognises several alternatives to this that may be 
more cost effective, such as the upgrade of the Ceduna antenna to 
L-band and/or the use of one of the AuScope geodetic antennas in WA 
at S-band.  All these options should be studied at greater length.
The addition of antennas to \tel\ beyond the current plans will make for a 
more sensitive VLBI array but will not materially open 
up new areas of science for VLBI.

\newpage
\section{Pulsar Science}\label{chap:psr}
{\bf Lead authors: {\em G. Hobbs, V. Kaspi}}\\
{\bf Contributing authors: {\em M. Bailes, S. Chatterjee, S. Johnston, R. Manchester, I. Stairs}}\\

\subsection{Summary}
Since the initial discovery of pulsars in Cambridge (Hewish et
al. 1968), radio telescopes world-wide have played a major part in the
discovery of new pulsars.  Pulsars have been used as tools to address
some of the most fundamental questions in basic physics, allowing
precision tests of general relativity, investigations into the
equation of state of ultradense matter and the behaviour of matter and
radiation in the highest magnetic fields known in the universe. The
commissioning of \tel\ will provide the transition between the use of
large single-dish instruments (such as the Parkes and Arecibo
telescopes) to using large numbers of small antennas with a wide
field-of-view (FoV) as is likely in the final mid to high frequency
SKA design.

Applications of pulsar observations include a very wide variety of
interesting astrophysical topics, often with unparallelled
precision.  Such topics include binary evolution, binary dynamics, the
interstellar medium, globular cluster physics, supernova remnant
astrophysics, the physics of relativistic winds and precision astrometry.
Pulsars are also instrinsically interesting, being the result of core collapse
supernovae and astonishing converters of mechanical energy of
rotation into electromagnetic radiation, particles and magnetic
fields.  The study of pulsars themselves is important for constraining
the overall population's properties and hence their origin, as well as
understanding the mysterious pulsar emission mechanism.

Historically it has been common to carry out pulsar research using
large single dish instruments.  \tel\ provides a pathway between the
current systems and the full-scale SKA.  \tel\ will be ideal for
carrying out relatively fast all-sky surveys that will allow $\sim
1000$ pulsars to be discovered.  In this chapter, we show that it will
be possible for \tel\ to continue to observe these (and previously
known) pulsars in order to determine their astrometric, spin and
orbital parameters.  \tel\ will, for several millisecond pulsars, be
able to provide data suitable for integration into a global timing
array project which aims to detect low-frequency gravitational wave
sources. 

Pulsar observations, especially searches, with \tel\ will present
several logistical and computational challenges, some of which can be
mitigated through specification choices.  In particular, any
large-scale survey with good sensitivity will be severely limited
computationally; the need to process every pixel
independently ensures that this will become worse as the square of the
maximum baseline length.  It is therefore necessary that as many short
baselines as possible be present in the configuration of \tel. For
pulsar timing, there is no such requirement; any configuration is
adequate.

The expansion of \tel\ would be good for several reasons.  Most
important is the increase in instantaneous sensitivity through both
the addition of more telescope collecting area and a reduction in
system temperature.  These will ensure that \tel\ will become more
competitive in searching and in high precision pulsar timing, leading
to improved data-sets for inclusion in global timing array projects.
More array elements will also ensure that more short baselines are
present in the array.

\newpage
\section{The Transient and Variable Radio Sky}\label{chap:trans}
{\bf Lead authors: {\em S. Johnston, I. Stairs}}\\ 
{\bf Contributing authors: {\em R. Ekers, C. James, J.-P. Macquart, R. Protheroe, B. Schmidt, M. Walker}}\\

\subsection{Summary}
The good sensitivity and large field-of-view
make \tel\ a unique instrument for studying radio transients and variables.
We strongly expect that \tel\ will make a unique discovery in the
transient parameter space, although it is naturally hard to
predict the nature of a new class of transients.

The key to a successful transient instrument is that it have high
sensitivity, large field-of-view, good dynamic range and high resolution.
\tel\ fulfills these criteria with its ability to achieve sub mJy
sensitivity across the entire sky in a single day observing.
Nearly all transients arise from point source objects; high resolution
is ideal for obtaining accurate positions necessary for follow-up
at other wavebands. It is important also that a wide range of timescales
from seconds to months are covered by the transient detector. This implies
a careful search strategy for uncovering rare objects.

\tel\ will most likely suffer from a surfeit of transient 
and variable sources.
This will pose challenges both for imaging and for determining which
sources are most interesting and worthy of follow-up observations
on other facilities. Chapter~\ref{chap:vlbi} describes some possible
mechanisms for VLBI follow-up of transient sources.

\newpage
\section{Summary}\label{chap:summ}
{\bf Section authors: {\em S. Johnston, R. Taylor, J. Wall}}\\

  Most of the phenomena we observe today, using telescopes to observe across the
  electromagnetic spectrum, were unknown a few decades ago. {\em Most were 
  discovered by radio astronomers using increasingly powerful instruments; see the
  list by Wilkinson et al. (2004).}

  In ``Cosmic Discovery'', Harwit (1981) addresses the question of
  what factors lead to new discoveries in astronomy. He argues that a large
  fraction of the discoveries have been associated with improved
  coverage of the electromagnetic spectrum or better resolution in the angle,
  time, or frequency domain. He also notes that astronomical discovery is often 
  closely linked to new technology introduced into the field. Wilkinson et~al. 
  (2004) detail just how it is that SKA will extend the
  multi-dimensional observing space, and they describe considerations in design and 
  operation that are important to enable discovery of the unexpected. These latter 
  precepts are equally vital for the design and operation of \tel.

  \tel\ is the first of a new generation of radio telescopes using
  innovative phased-array feed technology to explore a greatly enlarged
  survey-area parameter space.  This wide field of view -- a major step along the 
  SKA path -- is essential for many of the science cases presented here. But following 
  Harwit and Wilkinson et~al., we can anticipate it leading to the discovery of new, 
  rare and unexpected phenomena.
  As pointers to this, Section 2.9 describes one recent and 
  unexpected discovery, while Section 8.8 presents an
  unusual proposal which would open a new field of particle
  astrophysics to astronomers. One explores the capabilities of highest frequency resolution; the
  other the possibilities of using the highest time resolution.

  These exciting possibilities spotlight advancing radio astronomy: ASKAP is a key
  step on the strategic pathway towards the SKA. The goals of ASKAP, simply stated,
  are to carry out world class, ground-breaking observations, to demonstrate and 
  prototype technologies for the mid-frequency aspect of SKA, and to establish a
  site for radio astronomy in Western Australia where observations can be carried
  out free from the harmful effects of radio interference. 

  This paper has set out the main science themes to be tackled by ASKAP, themes which 
  play into the major issues confronting astrophysics and cosmology today and which
  naturally parallel those outlined as science drivers for the SKA: 
  the formation, evolution and population of galaxies including our own, 
  understanding the magnetic universe, exploring the poorly-understood transient 
  radio sky, and directly detecting the gravitational waves which must permeate our
  Universe.

\end{document}